\def\@cite#1#2{$^{\mbox{\scriptsize #1\if@tempswa , #2\fi}}$}

\def\@maketitle{%
  \newpage\spacing{1}\setlength{\parskip}{12pt}%
    {\Large\bfseries\noindent\sloppy \textsf{\@title} \par}%
    {\noindent\sloppy \@author}%
}

\newenvironment{affiliations}{%
    \setcounter{enumi}{1}%
    \setlength{\parindent}{0in}%
    \slshape\sloppy%
    \begin{list}{\upshape$^{\arabic{enumi}}$}{%
        \usecounter{enumi}%
        \setlength{\leftmargin}{0in}%
        \setlength{\topsep}{0in}%
        \setlength{\labelsep}{0in}%
        \setlength{\labelwidth}{0in}%
        \setlength{\listparindent}{0in}%
        \setlength{\itemsep}{0ex}%
        \setlength{\parsep}{0in}%
        }
    }{\end{list}\par\vspace{12pt}}

\documentclass[12pt]{article}
\usepackage{graphicx}
\usepackage{amssymb}
\usepackage[superscript,nomove]{cite}
\usepackage{times}
\usepackage{fullpage}

\large\normalsize 

\title{Planet heating prevents inward migration of planetary cores}
\author{Pablo Ben\'\i tez-Llambay$^1$, Fr\'ed\'eric Masset$^2$, Gloria Koenigsberger$^2$, and
  Judit Szul\'agyi$^3$}

\newcommand\aj{{Astron. J.}}%
\newcommand\araa{{Ann. Rev. Astron. Astrophys.}}%
\newcommand\apj{{Astrophys. J.}}%
\newcommand\apjs{{Astrophys. J. Suppl. Ser.}}%
\newcommand\aap{{Astron. Astrophys.}}%
\newcommand\aaps{{A\&AS}}%
\newcommand\mnras{{Mon. Not. R. Astron. Soc.}}%
\newcommand\nat{{Nature}}%
\newcommand\icarus{{Icarus}}%
\begin{document}

\maketitle

\parindent=0mm

\begin{affiliations}
\item IATE, Observatorio Astron\'omico, Universidad Nacional de 
  C\'ordoba, Laprida 854, C\'ordoba, X5000BGR, Argentina 
\item Instituto de Ciencias F\'\i sicas, Universidad Nacional 
  Aut\'onoma de M\'exico, Av. Universidad s/n, 62210 Cuernavaca, Morelos,
  Mexico 
\item University of Nice-Sophia Antipolis, CNRS, Observatoire de la C\^ote d'Azur, Laboratoire Lagrange, F-06304, Nice, France 
\end{affiliations}


{\bf Planetary systems are born in the disks of gas, dust and rocky
  fragments that surround newly formed stars. Solid content assembles
  into ever-larger rocky fragments that eventually become planetary
  embryos. These then continue their growth by accreting leftover
  material in the disc. Concurrently, tidal effects in the disc cause
  a radial drift in the embryo orbits, a process known as
  migration\cite{gt80,w97,2012ARA&A..50..211K,2013arXiv1312.4293B}.
  Fast inward migration is predicted by theory for embryos smaller
  than three to five Earth
  masses\cite{pbk11,2010ApJ...723.1393M,tanaka2002}.  With only
  inward migration, these embryos can only rarely become giant planets
  located at Earth's distance from the Sun and beyond\cite{2010AJ....139.1297L, 2014A&A...569A..56C}, in contrast
  with observations\cite{2013Sci...340..572H}. Here we report that asymmetries in the
  temperature rise associated with accreting infalling material\cite{1996Icar..124...62P,2014arXiv1406.5604M}
  produce a force (which gives rise to an effect that we call ``heating
  torque'') that counteracts inward migration. This provides a channel
  for the formation of giant planets\cite{2010AJ....139.1297L} and also explains the strong
  planet-metallicity correlation found between the incidence of giant
  planets and the heavy-element abundance of the host stars\cite{2005ApJ...622.1102F,2015AJ....149...14W}.}

We solve the equations governing the disc hydrodynamics in combination
with the equations of radiative transfer.  Planets have an angular
momentum that increases with their orbital radius. In the case of a
nearly circular orbit, the rate of change of angular momentum, or
torque, gives the migration rate. Our calculations are performed in
three dimensions, yielding a reliable value for the net torque, from
which the direction and rate of migration are inferred.

Our fiducial computation is one in which a rocky core with 3 Earth
masses is located at a distance comparable to that of Jupiter from the
Sun and is being bombarded by solid material at a rate that doubles
its mass in 100 thousand years.  We assume that the gravitational
energy of the infalling solid material is transformed entirely into
heat and ultimately radiated by the
planet~\cite{1996Icar..124...62P}. A second computation is performed
with the same set up, but without the planet's radiation, in order to
distinguish the effects of the heating torque from other torques.  We
find that the heating torque (defined as the torque difference between
cases with accretion turned respectively on and off) has a positive sign
(figure~1), which enables it to counteract the
effect of the standard, negative torque. The latter includes all
torque components of the non-heating case, and is always negative for
small mass embryos (typically smaller than $5\;M_\oplus$, where
$M_\oplus$ is the Earth's mass). Thus, the effect of the
heating torque is to either slow down the inward migration, cancel it,
or reverse its direction.

The most important factors governing the strength of the heating
torque and thus, the direction of migration, are the accretion rate of
the embryo, its mass and the opacity of the disc.  For our fiducial
values of opacity, disc structure and embryo mass, we find that
outward migration occurs for accretion rates corresponding to a mass
doubling time less than approximately 60~thousand years.  For larger
mass doubling times (i.e., smaller accretion rates), the heating
torque can at best slow down the inward migration but not reverse it
(figure~2).

 The heating torque has a large efficiency over the mass
interval $0.5-3\;M_\oplus$, which is precisely the range of masses
where counteracting inward migration is required in order to allow
further embryo growth at distances where giant planets are expected to
form~\cite{2014A&A...569A..56C}. Masses smaller than $0.5\;M_\oplus$,
for which the heating torque has a lower efficiency, migrate inward
only a negligible fraction of their orbital radius by the time they
double their mass.

Some insight into the physics of our new torque component can be
gained by examination of a mass density map in the planet vicinity.
The energy released by the planet heats the optically thick disc in
its vicinity. As the latter maintains its pressure equilibrium, hotter
regions are less dense than the surrounding ambient material.  As a consequence of the flow pattern around
the planet, two under-dense lobes appear, one leading and one
following the planet, when the heating is included in the calculation
(figure~3). Protoplanetary discs rotate in general at a
speed slightly smaller than the Keplerian speed, and the corotation
(the location at which the material orbits the star at same pace as
the planet) lies slightly toward its star from the planet.  This situation
favours the lobe that appears behind the planet: its material
approaches closer to the planet, receives more heat and is
consequently less dense than the other lobe, leading to a positive
torque on the planet. We explored the dependence of the heating torque
on the gradient of surface density (which leads to the shift between
the planet and its corotation), and we found that the heating torque
does indeed scale with the distance to corotation. Only in the
non-realistic cases in which the corotation is further out than the
planet (when the gas pressure increases outward) do we find a negative
heating torque. The heating torque therefore constitutes a robust trap
against inward migration in any realistic disc, when accretion rates
are large enough.

In the limit of a very large accretion rate, the heating torque
largely dominates other torque components, and the net torque is
directly proportional to the accretion rate. Since the migration rate
is proportional to the torque, it follows that the final distance over
which an embryo migrates is a function of the mass it accretes.  We
find that an embryo initially smaller than the Earth would at most
double its semi-major axis by the time it reaches $5$~Earth masses.


The finding that the heating torque can produce outward migration for
accreting embryos in their earliest stage of development has
implications for the formation of planetary systems in general, and
the Solar System in particular.  It opens a new route for the
formation of gas giant planets and alleviates the problem encountered
by current models of planetary population synthesis which predict too
many super-Earths~\cite{2014A&A...569A..56C} and a low yield of giant
planets.  The general picture that now emerges is that embryos with
masses in the range 0.3--5 Earth masses are able to avoid inward
migration when accretion rates are large.  By the time the heating
torque efficiency drops, they have entered a regime in which other
mechanisms driving outward migration come into
play~\cite{2014A&A...569A..56C,pbk11,2013A&A...549A.124B}.  Embryos
that are formed when accretion rates are low will still undergo inward
migration but at a slower rate. The overall migration  behaviour therefore displays a
bifurcation depending on the accretion rate of solids.

This bifurcation provides a simple and natural explanation for the
very strong correlation found between the incidence of giant planets
and the heavy-element content (i.e., metallicity) of the host
star~\cite{2005ApJ...622.1102F,2015AJ....149...14W}.  That is, since
the heating torque scales with the accretion rate and the accretion
rate, in turn, scales with the amount of solid content (a proxy of
which is the metallicity), protoplanetary discs with larger
metallicity will engender planets that can avoid inward migration and
grow to become giant planets. In contrast, embryos born in
lower metallicity environments cannot avoid inward migration, leading
to results as hitherto found in models of
planetary population synthesis, with low yields of giant planets and
ubiquitous super-Earths. These might therefore be more abundant in
metal-poor systems, as suggested by a recent planet search with
accurate stellar metallicities~\cite{2013EPJWC..4705001J}. The
incidence of super-Earths with the metallicity  of the host star is
however debated~\cite{2015AJ....149...14W}.

Recent models have contemplated the \emph{in situ} formation of
super-Earths at small orbital distances from the
star~\cite{2013ApJ...775...53H,2013MNRAS.431.3444C} ($\lesssim 1$~AU, one
AU being the average distance from the Earth to the Sun). We have not
performed calculations for embryos so close to the star, where the disk
parameters are still poorly constrained, but we note that the very
short mass doubling times inferred for some planets very close to
their star~\cite{2013MNRAS.431.3444C} would probably yield an extremely large
heating torque, the magnitude of which remains to be properly
calculated to assess the likelihood of \emph{in situ} formation scenarios.

The implications for the formation of our own Solar System are
somewhat more speculative, but are important to highlight.  The
parameters of our fiducial run are typical of those found beyond the
snow line (the distance at which water ice condenses). On the warm
side of the snow line we expect the heating torque to have a much
reduced efficiency, because the bombardment rate and the disc opacity
drop substantially.  With a heating torque producing outward
migration, we expect all embryos having formed beyond the ultimate
location of the snow line to have experienced a sizable outward
migration, thus causing a large depletion of solid material in this
region.  Hence, a prediction of our torque mechanism is that a
depleted region should be present inside the orbit of the first giant
planet in many planetary systems.  In our Solar System, such a region
may correspond to the asteroid belt, which only contains approximately
0.001 Earth mass of solid material~\cite{2011Natur.475..206W}.  Within
our framework, Jupiter's rocky core could have been assembled from
embryos originating from this region.

\emph{A priori}, the heating torque would not have been expected to
have the same order of magnitude as the tidal torque. It depends
primarily on the planet's physical radius, dust opacity and mass
accretion rate, whereas the tidal torque depends on the gaseous disc's
surface density and temperature. The serendipitous coincidence of both
torques is remarkable and allows the bifurcation between inward and
outward migration to occur for accretion rates that lie within the
range of the largest and smallest accretion rates believed to prevail
in protoplanetary systems.

Although further implications of this new torque component can only be
derived from a fine-tuned analysis of all parameters of the disc and
the embryo, it is clear that the heating torque provides a unifying
mechanism for gaining a deeper understanding of giant planet formation
in general and the specifics which gave rise to our own system.

\noindent{\bf Acknowledgements}

The authors wish to thank A. Morbidelli for a critical reading of a
first version of this manuscript. P.B.Ll. thanks CONICET for financial
support. This research was supported by UNAM grants PAPIIT IA101113
and IN105313 and by CONACyT grants 178377 and
129343. J.Sz. acknowledges support from the Capital Fund Management's
J.P. Aguilar Grant.
The authors wish to thank Ulises Amaya, Reyes 
Garc\'\i a  and J\'er\^ome Verleyen for their assistance in setting up the GPU cluster on which 
the calculations presented here have been run. 

\bigskip 

\noindent{\bf Contributions}
 
P.B.Ll. performed the numerical simulations and their subsequent
reduction.  F.M. designed the project and wrote the Methods
section. G.K. wrote the Main section. J.Sz. provided assistance with
the radiative transfer module. All authors contributed to the
discussion presented in this manuscript.

\bigskip 

\noindent{\bf Competing financial interests}

The authors declare no competing financial interests.

\bigskip 

\noindent{\bf Corresponding author}

Correspondence to: Fr\'ed\'eric Masset.

\bigskip

\let\oldthebibliography=\thebibliography 
\let\oldendthebibliography=\endthebibliography 
\renewenvironment{thebibliography}[1]{%
    \oldthebibliography{#1}%
    \setcounter{enumiv}{19}%
}{\oldendthebibliography}
%
%

\bigskip

\noindent{\bf Methods}

\bigskip

\noindent{\bf Code and numerical method}

We use the publicly available hydrocode FARGO3D ({\tt http://fargo.in2p3.fr}) to solve the hydrodynamics and radiative transfer equations on a spherical mesh, spanning in azimuth the whole range $[-\pi,\pi]$, in radius the range $[a/2,(3/2)a]$ (where $a$ is the planet's semi-major axis), and in colatitude the range $[\pi/2-3h,\pi/2]$ (where $h=H/r$ is the aspect ratio of the disc). The governing equations of the hydrodynamics module are the continuity equation, the Navier-Stokes equations including all components of the viscous stress tensor, and the energy equation. The equation of state of ideal gases is used. At each hydrodynamical time step, in addition, we solve the radiative energy on the mesh, using flux limited diffusion and a two temperature approach in the grey approximation, and we solve the coupling between thermal and radiative energies~\cite{2013A&A...549A.124B}.
These techniques are standard in the context of protoplanetary disks, in which they allow a simultaneous description of hydrodynamical and radiative effects at reasonable computational cost~\cite{2014arXiv1409.3011R}.

\noindent{\bf Fiducial calculation}

The parameters of our fiducial run are:
\begin{itemize}
\item A constant opacity $\kappa=1$~cm$^2$ g$^{-1}$ 
\item A kinematic viscosity $\nu=10^{15}$~cm$^2$ s$^{-1}$ 
\item A planetary mass $M_p=3$~M$_\oplus$ 
\item A planetary orbital radius $a=5.2$~AU 
\item The ratio of the specific heat at constant pressure to the
  specific heat at constant volume $\gamma =1.4$ 
\item The mean molecular weight of the gas  $\mu=2.3$~g mol$^{-1}$ 
\item A surface density law given by:
 \begin{equation}
   \label{eq:1}
   \Sigma(r) = 200\left(\frac{r}{a}\right)^{-\sigma}\mbox{~g cm$^{-2}$},
 \end{equation}
 with $\sigma=1/2$. This results in a value at $5.2$~AU that is
 roughly $30$~\% above the value quoted for the Minimum Mass Solar
 Nebula ~\cite{1985prpl.conf.1100H} (protoplanetary disk of minimum
 mass needed to form the Solar System).
\end{itemize}
The planetary potential is given by:
 \begin{equation}
   \label{eq:2}
   \phi_p=-\frac{GM_p}{(r^2+\epsilon^2)^{1/2}},
 \end{equation}
where $G$ is the gravitational constant, $r$ is the distance to the planet, and $\epsilon=2.5\cdot 10^{-3}a$ a softening length used to avoid a divergence of the
force in the planet vicinity.

\noindent{\bf Comparison with a second hydrocode}

The numerical scheme of the FARGO3D code is in may aspects similar to that of the ZEUS code~\cite{zeus}.  There are, however, two notable differences: FARGO3D conserves (angular) momentum to machine accuracy, and it features orbital advection (which gives its name to the code through the acronym of {\it Fast Advection in Rotating Gaseous Objects}~\cite{fargo2000}). 
We have checked our results on the fiducial run using a version of the nested mesh code JUPITER \cite{2014ApJ...782...65S}  that features a MUSCL-Hancock predictor step together with an exact adiabatic Riemann solver. As in FARGO3D, at each hydrodynamical time step, a radiative diffusion module based on flux limited diffusion in a two temperature approach is used to update the radiative and thermal energies. Solving for the radiative energy across the different mesh levels is done as in \cite{2014A&A...563A..11C}.
The heating torque that we find with this alternate code is within 25\% of the heating torque found with FARGO3D on the same setup.

\noindent{\bf Numerical setup}

Our mesh has resolution $1024\times512\times64$ in, respectively, azimuth, radius, and colatitude, with cell interfaces evenly spaced along each dimension.  Since the energy is released in the immediate vicinity of the planet, the effect on the torque appears on a short timescale (typically the dynamical timescale), hence 
meaningful results can be obtained with short term simulations.
We therefore run all our simulations over only $10$~orbits. We have checked nevertheless with one setup (not presented here) that the heating torque is constant over a  duration of $100$~orbits,  longer than the horseshoe libration timescale of our planets, from which we can discard the possibility that the heating torque could be a transient effect that would appear upon the insertion of the planet in the disc.  For each given setup, we run in succession:
\begin{itemize}
\item A meridian ($r,\theta$) two dimensional simulation to allow the disc to relax toward hydrostatic and radiative equilibrium.
\item A first three dimensional calculation in which a planet is inserted without releasing energy in the ambient disc,
which takes as initial condition the outcome of the previous run.
\item A second three dimensional calculation, which also takes as initial condition the outcome of the first run, in which we insert a planet of same mass as introduced in the previous run but which this time is allowed to release energy.
\end{itemize}

Our mesh is designed so that the planet is at the intersection between cell interfaces in azimuth, radius and colatitude. The planet therefore lies at the centre of an eight cell cube. The energy it releases is added evenly to these eight cells at each time step.  When integrating the torque exerted by the disc on the planet, we cut off the contribution of the cells that lie within half a Hill radius of the planet \cite{2009A&A...502..679C}. We find this cut off to have only a mild effect on the torque value.

The amount of energy released by the planet per unit time is:
\begin{equation}
 \label{eq:3}
 \dot E = \frac{GM\dot M}{R_p}=\frac{GM^2}{\tau R_p},
\end{equation}
where $\tau=M/\dot M$ is the planetary mass doubling time, and $R_p$ the physical radius of the planet, calculated assuming a density $\rho=3$~g cm$^{-3}$.  The mass doubling time in our fiducial run is $\tau=10^5$~yrs. 
This equation applies to embryos that are not surrounded by a dense envelope,  and it neglects the latent heat for vaporisation of the material (which is at most of order $10^{11}$~erg/g, whereas $GM_p/R_p\sim 10^{12}$~erg/g for the fiducial run).  We make no assumptions regarding the size of the solids that are 
impinging onto the core.  Although we might expect a vigorous heating 
torque when an embryo is subjected to the potentially extremely 
effective ''pebble accretion'' mechanism, further progress is needed in 
assessing the altitude at which pebbles deposit their energy and the 
ultimate core luminosity. However, regardless of the core luminosity that 
might arise from pebble accretion, a minimal value for the luminosity is 
set by the accretion of planetesimals, which leads to mass doubling 
times typical of those considered in our analysis.

\noindent{\bf Exploration of parameter space}

We have performed systematic explorations of the magnitude of the heating torque by varying one parameter at a time of our fiducial set up. Namely, we have varied the viscosity between $4\cdot 10^{14}$ and $4\cdot 10^{15}$~cm$^2$ s$^{-1}$, the surface density between $100$ and $1000$~g cm$^{-2}$, the mass doubling time between $3\cdot 10^4$ and $3\cdot 10^5$~years, the opacity between $0.1$ and $10$ cm$^2$ g$^{-1}$, the planetary mass from one third of the Earth's mass to $7$ Earth's masses, and the exponent $\sigma$ of the power law of the surface density, from $-2$ to $1.5$.
These explorations show that the heating torque, normalised to the torque of the non-accreting case, depends only weakly on the surface density and viscosity.  The dependency on $\sigma$ is as mentioned in the main text: the heating torque is found to scale with the distance between the planet and corotation.  For the other parameters, we find the dependencies given in the Figure~4. The torque values reported in this figure are time-averages of the torque measured over orbits 5 to 10 of each calculation.

The heating torque depends strongly on the opacity and on the mass doubling time. At larger opacity, the energy radiated by the planet is trapped nearby it, hence the under-dense lobes are more pronounced and the net effect is stronger than for smaller opacity. When the mass doubling time decreases, the heating torque increases since $\dot E$ increases. We also find that the heating torque has a large efficiency over the mass interval $0.5-3\;M_\oplus$. Its decline past this mass is compatible with the fact that the distance of the planet to corotation becomes an ever smaller fraction of the Hill sphere, so that the flow in the vicinity of the planet (and therefore the under-dense lobes) become more symmetric.

A final comment is warranted  regarding the torque resulting from the angular momentum that is transferred 
to the planet by the material being accreted. We find it to be several orders of magnitude smaller than the 
heating torque, so it can be safely neglected. 

\noindent{\bf Code availability}

The FARGO3D code is publicly available at the address {\tt http://fargo.in2p3.fr}. This public version does not yet include the radiative transfer module, which will be made available once it has been fully documented. The JUPITER code is not publicly available at the present time.

\begin{figure}[ht!]
  \centering 
\includegraphics[width=\columnwidth]{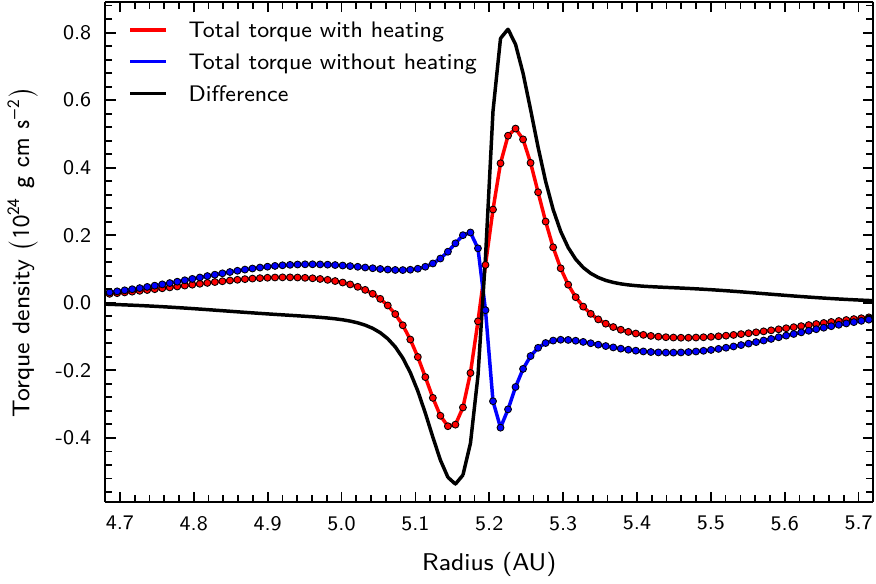}
\caption{Comparison of the torques in the cases with and without 
    heating. The blue curve shows the torque radial density (i.e. torque exerted by rings of unitary radial 
    width upon the planet) in 
    the non-heating case and the red curve when the heating is included. 
    Their difference shows the heating torque density 
    (black).  This calculation 
    corresponds to an embryo planet of 3~Earth masses which is located 
    at 5.2 AU from its central star. 
    \label{fig:tqraddens}}
\end{figure}

\clearpage

\begin{figure}[ht!]
  \centering 
  \includegraphics[width=.9\columnwidth]{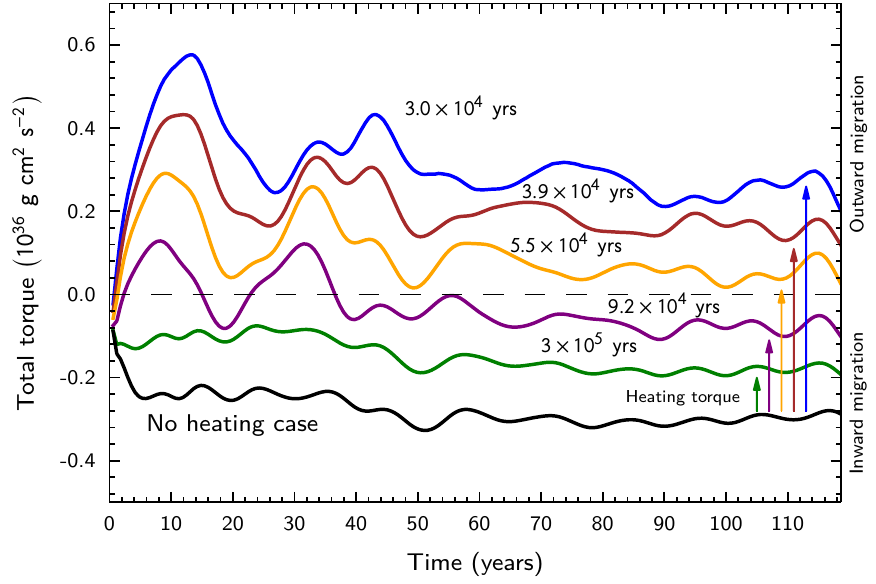}
\caption{Heating torque for different growth time scales. 
    The curves are labelled with 
    the planetary mass doubling time (i.e. the time it takes the 
    accreting planet to double its mass) and show the torque exerted on a 3 Earth masses 
    embryo planet over the first 118 years of our calculation (10 
    orbits). With  low or no heating, the planet migrates inward while for 
    larger rates (mass doubling time shorter than 92 thousand years) 
    it migrates outward. The dashed line 
    corresponds to no migration. The vertical arrows show the 
    magnitude of the heating torque. 
    \label{fig:tqvst}}
\end{figure}

\clearpage 

\begin{figure}[h!]
  \centering 
  \includegraphics[width=.8\columnwidth]{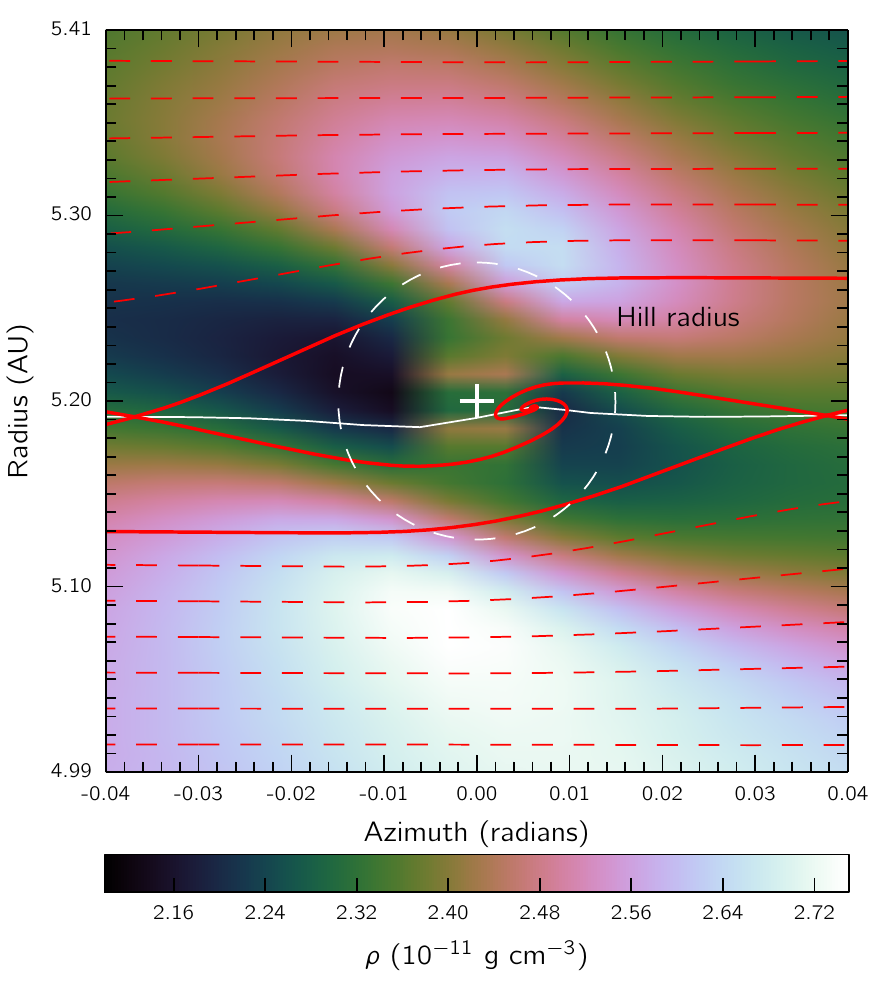}
\caption{Density in the vicinity of an    irradiating embryo. This equatorial slice of the mass 
    density shows 
    two low density lobes on each 
    side of the planet (cross). The more apparent one (left) gives 
    rise to the positive torque. Such lobes are absent 
    for a non-radiating embryo. The dashed 
    circle shows the planetary Hill radius. Streamlines are in 
    red showing separatrices of the coorbital region (bold) and 
 paths of material further from the planet (dashed). The white nearly horizontal curve shows 
    corotation, i.e. the place where the material is at rest in the 
    planet frame. 
    \label{fig:lobes}}
\end{figure}

\clearpage

\begin{figure}[ht!]
  \centering 
\includegraphics[width=.9\textwidth]{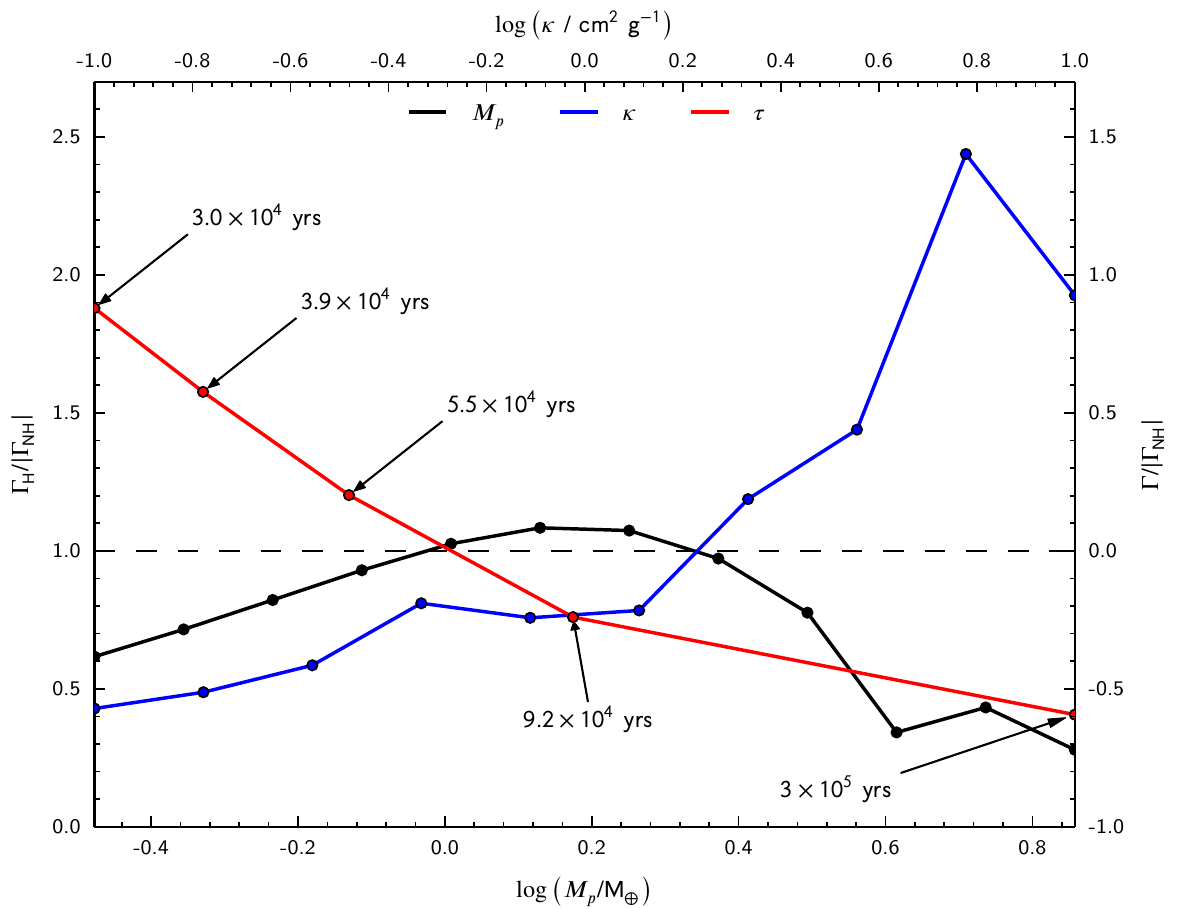}
\caption{Exploration of the parameter space. Heating torque $\Gamma_\mathrm{H}$ normalised to the absolute value of the torque of the non-accreting case $|\Gamma_{\mathrm{NH}}|$, as a function of embryo mass $M_p$, opacity $\kappa$ and mass doubling time $\tau$. Whenever one parameter is varying, others have the value of the fiducial run. Mass doubling times are given in units of years and show that a positive torque results for $\tau \lesssim 60,000$~years. The right axis shows the total torque $\Gamma=\Gamma_\mathrm{H}+\Gamma_\mathrm{NH}$, also normalised to $|\Gamma_{\mathrm{NH}}|$. The horizontal dashed line corresponds to no migration.}
\end{figure}

\end{document}